\documentclass[
aps,prl,
reprint,
superscriptaddress,
amsmath,
amssymb
]{revtex4-1}

\usepackage[utf8]{inputenc}
\usepackage{times}
\usepackage{microtype}
\usepackage{graphicx}
\usepackage{upgreek}
\usepackage{bm}
\usepackage{hyperref}
\usepackage{xcolor}
\usepackage[normalem]{ulem}
\usepackage{placeins}
\usepackage{dcolumn}

\usepackage{filecontents}
\usepackage[T1]{fontenc}
\usepackage{csquotes}
\usepackage[german,ngerman,english]{babel}
\usepackage{centernot}
\usepackage{mathtools}
\usepackage{ifthen}
\usepackage{tikz}
\usetikzlibrary{calc}
\usetikzlibrary{decorations.pathreplacing}
\usetikzlibrary{decorations}
\usetikzlibrary{positioning}
\usepackage{pgfplots}
\usepackage{xxcolor}
\definecolor{structure}{rgb}{0.23,0.4,0.7}
\usepackage[normalem]{ulem}

\newtheorem{theorem}{Theorem}

\newtheorem{conjecture}{Conjecture}

\newsavebox{\blocksavebox}

\definecolor{niceblue}{rgb}{0.33,0.5,0.8}

\newcommand{\refsub}[2]{\hyperref[#1]{\ref*{#1}#2}}

\newcommand{\eqqcolon}{\mathrel{=\mkern-1.2mu\vcentcolon}}

\renewcommand{\max}{\mathchoice{\operatorname*{max}}{\operatorname*{max}}{\mathrm{max}}{\mathrm{max}}} 
\renewcommand{\min}{\mathchoice{\operatorname*{min}}{\operatorname*{min}}{\mathrm{min}}{\mathrm{min}}}

\newcommand{\bra}[1]{\langle #1|}
\newcommand{\ket}[1]{|#1\rangle}

\newcommand{\abs}[1]{|{#1}|}
\newcommand{\norm}[2][]{
  \ifthenelse{\equal{#1}{}}
    {\left\| {#2} \right\|}
    {\ifthenelse{\equal{#1}{uinv}}
      {\left\vert\kern-0.25ex\left\vert\kern-0.25ex\left\vert {#2} \right\vert\kern-0.25ex\right\vert\kern-0.25ex\right\vert}
      {\left\| {#2} \right\|_{#1}}
    }
}

\newcommand{\taverage}[2][]{
  \ifthenelse{\equal{#1}{}}
  {\overline{#2}}
  {\overline{#2}^{#1}}
}

\newcommand{\tracedistance}[3][]{
  \ifthenelse{\equal{#2}{}}
  {\ifthenelse{\equal{#3}{}}
    {\mathcal{D}_{#1}}{}
  }{
    \ifthenelse{\equal{#1}{}}
    {\mathchoice{\operatorname{\mathcal{D}}\left(#2,#3\right)}{\operatorname{\mathcal{D}}(#2,#3)}{\operatorname{\mathcal{D}}(#2,#3)}{\operatorname{\mathcal{D}}(#2,#3)}}
    {\mathchoice{\operatorname{\mathcal{D}}_{#1}\left(#2,#3\right)}{\operatorname{\mathcal{D}}_{#1}(#2,#3)}{\operatorname{\mathcal{D}}_{#1}(#2,#3)}{\operatorname{\mathcal{D}}_{#1}(#2,#3)}}
  }
}
\newcommand{\fidelity}[3][]{
  \ifthenelse{\equal{#2}{}}
  {\ifthenelse{\equal{#3}{}}
    {\mathcal{F}_{#1}}{}
  }{
    \ifthenelse{\equal{#1}{}}
    {\mathchoice{\operatorname{\mathcal{F}}\left(#2,#3\right)}{\operatorname{\mathcal{F}}(#2,#3)}{\operatorname{\mathcal{F}}(#2,#3)}{\operatorname{\mathcal{F}}(#2,#3)}}
    {\mathchoice{\operatorname{\mathcal{F}}_{#1}\left(#2,#3\right)}{\operatorname{\mathcal{F}}_{#1}(#2,#3)}{\operatorname{\mathcal{F}}_{#1}(#2,#3)}{\operatorname{\mathcal{F}}_{#1}(#2,#3)}}
  }
}

\newcommand{\Sr}[3][]{
  \ifthenelse{\equal{#1}{}}
    {\operatorname{\mathnormal{S}}(#2\|#3)}
    {\operatorname{\mathnormal{S}}_{#1}(#2\|#3)}
}

\DeclareMathOperator{\1}{\mathds{1}}
\DeclareMathOperator{\id}{\mathrm{id}}



\newcommand{\mb}[1]{\mathbb{#1}}

\newcommand{\C}{\mb{C}} 

\newcommand{\Gap}{\Delta}
\newcommand{\GapParent}{\Delta_*} 
\newcommand{\DimPhysical}{d}
\newcommand{\DimVirtual}{D}
\newcommand{\DimSpatial}{\mathsf{d}}
\newcommand{\ErrorObservableTrueGSToPEPS}{\epsilon}
\newcommand{\ErrorObservablePEPSToPatch}{\varepsilon}
\newcommand{\ErrorTraceDistanceTrueGSToPEPS}{{\epsilon}}

\definecolor{jens}{rgb}{0.1,0.5,0.1}
\definecolor{martin}{rgb}{0,0,1.0}

\newcommand{\sharpp}{{$\mathsf{\#P}$}}
\newcommand{\pp}{{\sf PP}}
\newcommand{\qma}{{\sf QMA}}
\newcommand{\postbqp}{{\sf PostBQP}}

\newcommand{\beq}[0]{\begin{equation}}
\newcommand{\eeq}[0]{\end{equation}}
\newcommand{\poly}{\mathrm{poly}}
\newcommand{\hide}[1]{}

\begin{document}

\title{Approximating local observables on projected entangled pair states}
\author{M. Schwarz, O. Buerschaper, and J. Eisert}
\address{Dahlem Center for Complex Quantum Systems, Freie Universit{\"a}t Berlin, 14195 Berlin, Germany}

\begin{abstract}
Tensor network states are for good reasons believed to capture ground states of gapped local Hamiltonians arising in the condensed matter context, states which are in turn expected to satisfy an entanglement area law. However, the computational hardness of contracting projected entangled pair states in two and higher dimensional systems 
is often seen as a significant obstacle when devising higher-dimensional variants of the density-matrix renormalisation group method. In this work, we show that for those projected entangled pair states that are expected to provide good approximations of such ground states of local Hamiltonians, one can compute local expectation values in quasi-polynomial time. We therefore provide a complexity-theoretic justification of why state-of-the-art numerical tools work so well in practice. We comment on how the transfer operators of such projected entangled pair states have a gap and discuss notions of local topological quantum order. 
We finally turn to the computation of local expectation values on quantum computers, 
providing a meaningful application for a small-scale quantum computer.
\end{abstract}

\maketitle

Recent years have seen an explosion of interest in capturing quantum many-body systems in terms of tensor network states 
\cite{OrusReview,VerstraeteBig,EisertReview,Schollwock201196}. Such approaches provide powerful numerical tools
for simulating strongly correlated quantum systems \cite{PEPSKagome,VerstraeteBig,iPEPSOld,iPEPS,Lubasch}, 
even for fermionic systems \cite{MERAF1,MERAF2,Corboz2DHubbard,MERAF4,CorbozPEPSFermions}, 
overcoming the notorious sign problem that marres Monte Carlo approaches. The success of such approaches is essentially rooted in 
the entanglement structure that ground states of gapped local Hamiltonian models exhibit: They are expected to satisfy an 
entanglement area law \cite{AreaReview}, originating from the locality of interactions.
The insight that ground states are very a-typical quantum states is often captured in the phrase that states having this entanglement pattern constitute what is called the 
``physical corner'' of Hilbert space \cite{poulin2011quantum}. Indeed, the intuition that tensor network states should approximate this physical corner remarkably well is substantiated by a significant body of numerical work. In this discussion,  \emph{projected entangled pair states (PEPS)}
\cite{PEPSOld,PEPSKagome,iPEPSOld,iPEPS,Lubasch},
the higher-dimensional analogues of \emph{matrix product states} (MPS), take the leading role. Such PEPS not only provide numerical tools, but are also workhorses 
for understanding phases of matter or  notions of topological order \cite{ClassificationPhases,Buerschaper-AnnPhys-2014,PEPSTopology}.

This development can actually be seen as a natural continuation of the established \emph{density-matrix renormalisation group (DMRG)} 
method that allows to simulate 1D quantum systems essentially to machine precision \cite{White,Schollwock201196}. For such 1D systems, a deep understanding
on the functioning of tensor networks has already been reached, even to full rigor. Area laws for entanglement entropies have been 
proven to hold for gapped models  \cite{OneD}, implying MPS approximations \cite{SchuchApprox}. A
polynomial-time classical algorithm computing an MPS approximation of ground states of gapped Hamiltonians
\cite{Vidick1D} can be seen as a DMRG method with a convergence proof, at the cost of less efficiency.

But even for higher dimensional systems, the same intuition is expected to be valid. Strictly speaking, area laws alone may not be sufficient to guarantee 
that PEPS are good approximations of given quantum states \cite{1411.2995}. But the expectation that the physical corner is well approximated by 
PEPS, dubbed the ``PEPS conjecture'', is still very much in place: This expectation is usually taken for granted and constitutes the basis of an entire research field. Indeed, 
for higher temperatures,  a variant of this conjecture is actually provably true \cite{ThermalPEPOs,Intensive}.

Having said all this, a new obstacle emerges for higher-dimensional systems; one that is often seen as a key obstacle, a make-or-break issue when it comes to
numerically simulating strongly correlated models with PEPS: Even though PEPS are expected to provide good approximations, they are believed to be 
not efficiently contractible to compute expectation values of local observables. This is backed up by a proof in worst-case complexity, stating that the contraction of
two-dimensional PEPS is \sharpp-complete \cite{Contraction}. This is seen as a key burden for further developing such numerical tools. It creates a somewhat ironic situation that while the right variational principle has been identified, it may well be that one cannot compute the relevant features efficiently. This observation is also to some extent at odds with
a large body of numerical evidence \cite{PEPSKagome,iPEPSOld,iPEPS,Lubasch,Corboz2DHubbard}, showing that in practice, this hardness of 
contraction is not so much of a problem. Rather PEPS contraction gives rise to reliable results. This is even more so the case when one starts with an exact
PEPS construction in the first place \cite{ResonatingValenceBond}.

In this work, we contribute to clarifying this dichotomy. We show that while general PEPS may well be computationally hard to contract, this does not apply to the same extent to those PEPS that are expected to provide good approximations to ground states. Frankly put -- and made more precise below -- we arrive at the following situation: Either a PEPS is a
good approximation of a given ground state, and then a (quasi-) polynomial time contraction is perfectly possible. Or it is not, but then the issue of contracting PEPS does not arise anyway.  We will make this notion precise in the following argument. 

\emph{PEPS conjecture.}
\emph{Tensor networks} and in particular \emph{projected entangled pair states (PEPS)} are generally expected to describe ground states of gapped 
many-body models exceedingly well. Recent years of numerical and analytical work on tensor network states have largely clarified that the ``physical corner'' of 
Hilbert space may indeed be well parametrised by PEPS. Still, interestingly, a  ``PEPS conjecture'' that specifies 
in what precise way one expects PEPS to provide good approximations has not yet been formulated in written form, albeit being common knowledge in the community. 
Here we present two readings of what could reasonably be called a PEPS conjecture.
We consider local Hamiltonians $H=\sum_i h_i$ defined on a $\DimSpatial$-dimensional regular
lattice of spins with a \emph{constant} spectral gap $\Gap>0$ above the \emph{unique} ground state~$\rho$.
Furthermore we are interested in arbitrary observables~$O_X$ as long as the number of sites in their support~$X$ is upper bounded by a \emph{constant}~$k$.
In particular, observables supported on multiple disconnected regions are perfectly covered, e.g. two-point correlation functions.
The first conjecture merely states that PEPS are good approximations for ground states of gapped models. 

\begin{conjecture}[Weak PEPS conjecture] \label{conj:weakPEPS}
For all~$O_X$ and any  $\ErrorObservableTrueGSToPEPS>0$ there exists a PEPS~$\omega$
with bond dimension~$\DimVirtual=O\bigl(\poly(N,\ErrorObservableTrueGSToPEPS^{-1})\bigr)$ such that
$|{\rm tr}(O_X \rho)-  {\rm tr}(O_X \omega)|<\ErrorObservableTrueGSToPEPS$.
\end{conjecture}

This is provably true for 1D systems \cite{OneD,SchuchApprox,Vidick1D}, even in the stronger incarnation that such a
matrix product state (MPS) approximation exists satisfying $\|\omega-\rho\|_1<\ErrorTraceDistanceTrueGSToPEPS$, implying the above. Similarly,
one can ask the PEPS $\omega$ to approximate the ground state $\rho$ in relative entropy 
$S(\omega\|\rho) \leq \ErrorTraceDistanceTrueGSToPEPS^2/2$, 
which implies the above by virtue of Pinsker's inequality 
$\|\omega-\rho\|_1\leq (2 S(\omega\|\rho))^{1/2}$.
For systems with $\DimSpatial>1$, the closest result to Conjecture \ref{conj:weakPEPS}
we are aware of is the one presented in ref.\ \cite{ThermalPEPOs}, which uses a
specific assumption on the density of states to find $\DimVirtual=e^{O(\log_2(N/\ErrorTraceDistanceTrueGSToPEPS)^{\DimSpatial+1})}$, 
which is quasi-polynomial in $N/\ErrorTraceDistanceTrueGSToPEPS$ for constant $\DimSpatial$. 
While this is perfectly reasonable, the conjecture misses the point that it does not necessarily capture key properties of the true ground state. 
Again for 1D systems, injectivity of the MPS  will readily imply exponentially decaying correlations. This excludes
situations of the kind where a state of the form $\lambda \rho+ (1-\lambda)\eta$ for $\lambda \geq 1-\ErrorTraceDistanceTrueGSToPEPS/2$ would provide an 
approximation for the ground state $\rho$ in the above sense, even if the latter state $\eta$ 
has correlation properties very different from those of $\rho$. In a stronger formulation of the PEPS conjecture, one 
would exclude such situations, arguing that it is in the nature of a good approximation to also reproduce the qualitative
behavior of a property such as the decay of correlations, in addition to providing  mere numerical 
numbers of local expectation values. One can hence 
argue that reasonable approximations of ground states of gapped models should themselves be ground states of gapped models.

\begin{conjecture}[Strong PEPS conjecture] \label{conj:strongPEPSconj}
For all~$O_X$ and any constant $\ErrorObservableTrueGSToPEPS>0$
there exists an \emph{injective} PEPS~$\omega$ with bond dimension~$\DimVirtual=O\bigl(\poly(N,\ErrorObservableTrueGSToPEPS^{-1})\bigr)$  
such that its parent Hamiltonian~$H_*$ has a constant spectral gap $\GapParent>0$ and $|{\rm tr}(O_X \rho)-  {\rm tr}(O_X \omega)|<\ErrorObservableTrueGSToPEPS$. 
\end{conjecture}

Importantly, this conjecture is again provably true even in higher dimensions for all states that are in the same phase as a product state -- so which are the 
 \emph{trivial phase}, viewed from a perspective of topological order: The relevant states can then be quasi-adiabatically prepared \cite{PhasesNachtergaele} from products, 
giving rise to a short ranged quantum circuit.   None of the above capture non-injective PEPS or non-unique
ground states, so that the above formulations do not capture intrinsic topological order. 
Following previous work \cite{PEPSgen1,PEPSgen2,ge2016rapid}, we restrict ourselves to the subset of models exhibiting a \emph{uniform} spectral gap, i.e. we ask for a constant spectral gap lower bound for \emph{a family} of parent Hamiltonians related to the parent Hamiltonian of the input PEPS. 
This is a natural and common feature for gapped local models, and can be proven to hold for classes of PEPS. At the same time, 
this property does not directly follow from the PEPS conjectures as both examples as well as counter-examples are known.

\emph{Main result.}
We now turn to showing how expectation values of local observables can be (quasi-)efficiently approximated on PEPS which are the ground states of local Hamiltonians. We will assume constituents referred to as ``spins'' to have finite dimension~$\DimPhysical$. We discuss  contraction properties of PEPS in general terms, which can of course be seen as PEPS that approximate ground states via Conjecture~\ref{conj:strongPEPSconj}.

\begin{figure}
 \begin{center}
 \includegraphics[width = .59\columnwidth]{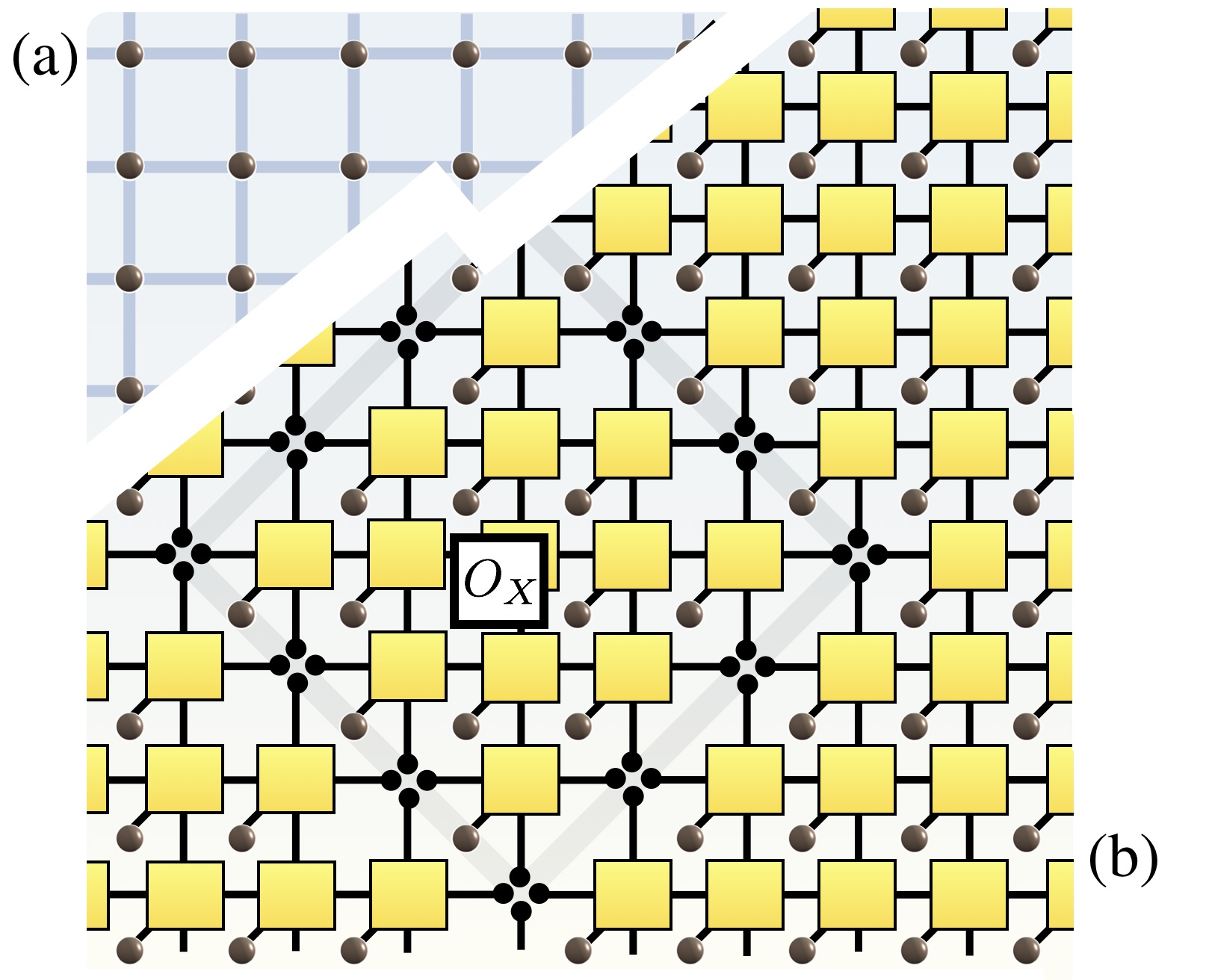}
 \end{center}
\caption{ \label{fig:Kita}
(a) A two-dimensional quantum lattice model. If the model is gapped, its ground state is conjectured to be well-approximated by a PEPS.
(b) PEPS tensors mapping \emph{virtual} to \emph{physical} indices.}
\end{figure}

\begin{theorem}[Computation of observables] \label{thm:PEPSobservable}
Let $\omega$ be an unnormalised, injective PEPS defined on a (constant) $\DimSpatial$-dimensional lattice of $N$ spins with bond dimension~$\DimVirtual$, and physical dimension~$\DimPhysical$,
such that its parent Hamiltonian $H_*$ is uniformly gapped with constant spectral gap $\GapParent>0$.  
Let $\{A_i\}$ be the collection of 
local tensors specifying the PEPS, and let $\kappa_*=\max_i \kappa(A_i)$ be an upper bound on the condition number. Let $O_X$ be  an
observable supported on $|X|<k$ sites for constant $k$. Then an approximation $\widetilde{O_X}$ of the expectation value $\bra{\omega}O_X\ket{\omega}/\langle\omega\vert\omega\rangle$ such that 
\begin{equation*}
\Bigl|{\bra{\omega}O_X\ket{\omega}}/{\langle\omega\vert\omega\rangle}- \widetilde{O_X}\Bigr| \leq\ErrorObservablePEPSToPatch,
\end{equation*} 
can be computed in 
 time $(\DimVirtual\DimPhysical)^{O(\ell^{\DimSpatial})}$ on a deterministic classical computer, and
time $\tilde{O}(\ell^\DimSpatial/\ErrorObservablePEPSToPatch^{2})$ and $O(\text{polylog}(\ell/\varepsilon))$ depth on a quantum computer,
where 
\begin{equation}\label{ell}
	\ell \in O\left(\frac{2 \ln(\kappa_*) + \ln(\ErrorObservablePEPSToPatch^{-1}) + \ln(\norm{O_X})}{\GapParent}\right).
\end{equation}
\end{theorem}
That is, the computation is possible in quasi-polynomial deterministic time or poly-logarithmic quantum depth for polynomially scaling $\ErrorObservablePEPSToPatch^{-1}$ and $\kappa_*$, and constant 
$\DimSpatial,\GapParent>0$. 
Or in poly-logarithmic quantum time for polynomial $\kappa_*$ and poly-logarithmic $\ErrorObservablePEPSToPatch^{-1}$ and constant 
$\DimSpatial,\GapParent>0$. 
Or in constant deterministic time for constant $\DimSpatial, \GapParent, \kappa_*, \ErrorObservablePEPSToPatch>0$. In the special case of $\DimSpatial=1$, the run-time of the algorithm scales polynomially in the 
system size, as expected for MPS. 
Note that the set $X$ does not have to be simply connected, so that our theorem
covers observables $O_X$ reflecting correlation functions. 
A related result has also been shown in ref.\ \cite{cirac2013robustness} assuming 
\emph{local topological quantum order (LTQO)}, a condition implying exponential decay of correlations.
Our result only uses the exponential clustering theorem -- which is provably true in all finite dimensions \cite{hastings2004lieb,math-ph/0506030} -- while 
LTQO is not known to hold in this generality.

\emph{Preliminaries.}
A \emph{tensor} of rank $r$ with dimension $D$ is a linear object in $(\C^{D})^{\otimes r}$, 
a $D^r$-dimensional array of complex numbers. A tensor $T_{i_1,\dots, i_r}$ with indices $i_s$, $1 \leq s \leq r$, that take values in the range $1 \leq i_s \leq\DimPhysical$, may represent amplitudes of a quantum state by associating the $r$ indices to $r$ 
physical qudits, i.e.~$\ket{\varphi}=\sum_{i_1,\dots,i_r=1}^\DimPhysical T_{i_1,\dots, i_r} \ket{i_1}\otimes \cdots \otimes\ket{i_r}$. Two rank $r$ tensors $S,T$ can be \emph{contracted along an index $j$} to form a rank $2r-2$ tensor $U$ by summing over the joint index. 
A contracted index is called \emph{closed}, whereas an uncontracted index is \emph{open}. 
We consider here \emph{PEPS tensor networks} based on cubic lattices 
of dimension~$\DimSpatial$, where vertices $v\in V$ are associated with tensors $T_v$. Each edge $e\in E$ between 
nearest neighbors connects indices of two tensors with matching bond dimension~$\DimVirtual$, called  \emph{virtual indices}. Furthermore, each tensor $T_v$ 
has an  open index of dimension~$\DimPhysical$, the \emph{physical index}.
Consequently, a PEPS defines a quantum state vector 
\begin{equation*}
\ket{\psi} = \sum_{i_1,\dots, i_N=1}^\DimPhysical \mathcal{C(T}_{i_1,\dots,i_N}\mathcal{)} \ket{i_1}\otimes \cdots \otimes \ket{i_N}
\end{equation*}
where the contraction is over all closed indices for each $i_1,\dots,i_N$,
according to the  \emph{contraction value}
$\mathcal{C(T)} = \sum_{i_e=1,\,e \in E}^\DimVirtual \prod_{v \in V} T_{v;i_e}$.
The term ``projected entangled pair state'' (PEPS) derives from the following alternative view of a tensor network: Put a \emph{maximally entangled pair} of qudits along each edge $e \in E$, i.e. $\ket{\phi_e} = \DimVirtual^{-1/2} \sum_{i=1}^\DimVirtual \ket{i, i}$. Then the tensors $T_v$ can be viewed as linear maps from the $r$ inbound virtual indices of dimension~$\DimVirtual$ at each vertex $v$ to the single physical index of dimension~$\DimPhysical$ called the PEPS \enquote{projector}
\begin{equation*}
A_v = \sum_{i=1}^\DimPhysical \sum_{j_1,\dots,j_r=1}^\DimVirtual T_{v;i,j_1,\dots,j_r} \ket{i}\bra{j_1,\dots,j_r}.
\end{equation*}
Using this map, the state vector of the PEPS can be written as
\begin{equation} 
\ket{\psi} = \bigotimes_{v \in V} A_v  \bigotimes_{e \in E} \ket{\phi_e} .
\label{eq:PEPSmap}
\end{equation}
A PEPS is called \emph{injective} \cite{PEPS,Perez-Garcia-QIC-2008} iff each PEPS projector $A_v$ has a left-inverse, i.e. $A_v^{-1}A_v=\id$, where $A_v^{-1}$ is the Moore-Penrose inverse. The injectivity condition also holds, if the left-inverse exists only after
\emph{blocking} of a constant number of tensors, merging a constant number of adjacent physical indices in turn. On a $\DimSpatial$-dimensional lattice, it is also clear from dimension counting alone that, after blocking, the PEPS projector maps the virtual space of smaller dimension to the physical space of larger dimension.
Any non-injective PEPS is $\ErrorTraceDistanceTrueGSToPEPS$-close to an injective one.
The main result about injective PEPS is a standard construction of a frustration-free local \emph{parent Hamiltonian}, which has this PEPS as its unique ground state 
\cite{Perez-Garcia-QIC-2008}.
W.l.o.g., let us thus assume in this paper that the injective PEPS under consideration is already blocked, such that individual tensors $A_v$ are left-invertible.
Consequently, we may assume that the parent Hamiltonian has interactions between nearest neighbours only. We say a PEPS defined by local tensors $\{A_v\}_{0 \leq v\leq N}$ has a \emph{uniformly gapped} parent Hamiltonian $H_*$ with spectral gap $\Delta_*$, if every member $H_t$ of the family $\{H_t\}_{0 \leq t \leq N}$ of parent Hamiltonian of the sub-PEPS $\{A_v\}_{0\leq v \leq t}$ has a spectral gap $\Delta_t \geq \Delta_*$ \cite{PEPSgen1,PEPSgen2,ge2016rapid}.

\emph{Proof sketch of Theorem 1.} We present here the core argument of the proof, with further details stated in the supplementary material.
The key idea is to disentangle a boundary of size $O(\ell^{\DimSpatial-1})$ of PEPS tensors $\{A_{i}\}$ around the observable~$O_X$ from the PEPS step by step, while bounding the additive error  in estimating the expectation value of~$O_X$ introduced in each step.
In this way, we achieve an overall error bound that scales linearly with the number of tensors removed, that is $O(\ell^{\DimSpatial-1})$ times the per-step error. A key tool we use is the \emph{exponential clustering theorem}, which is known to hold in all finite dimension \cite{hastings2004lieb,math-ph/0506030}. To some extent, our approach can also be viewed as partially inverting the PEPS preparation algorithms as discussed in Refs.\ \cite{PEPSgen1, PEPSgen2, ge2016rapid}.
For an unnormalised, injective PEPS~$\ket{\omega}$, specified by the collection 
of tensors $\{A_i\}$, with a gapped, local parent Hamiltonian $H_*$ \cite{Perez-Garcia-QIC-2008},
we define a sequence of parent Hamiltonians $H_{i}$,  $0 \leq i \leq N$, 
on the same $\DimSpatial$-dimensional lattice, with formally \emph{normalised} ground states 
\begin{equation*}
	\ket{\omega_{i}} = \frac{A_{i} \cdot \dots \cdot A_1 \ket{\phi}^{\otimes n}}{\norm{A_{i} 
	\cdot \dots \cdot A_1 \ket{\phi}^{\otimes n}}},
\end{equation*}
so that $H_N=H_*$, $n=O(N)$ denoting the number of edges.
We assume that the last $N_b=O(\ell^{\DimSpatial-1})$ tensors in the sequence $(A_i)_{0\leq i \leq N}$ constitute a boundary at graph theoretical distance $\ell$ on the lattice around observable~$O_X$ (this 
sequence can always be constructed for an injective PEPS \cite{PEPS,Perez-Garcia-QIC-2008}). By assumption $H_*$ is uniformly gapped by $\Delta_*$, therefore $\Gap(H_{i}) \geq \Delta_*$ for all $H_i$, following ref.\ \cite{PEPSgen1,PEPSgen2,ge2016rapid}.
%
The exponential clustering theorem \cite{hastings2004lieb,math-ph/0506030} applies to each $H_i$, such that for each $i$ and fixed $O_X$
\begin{align}
\left |\bra{\omega_{i}} O_X \otimes O_i \ket{\omega_{i}} - \bra{\omega_{i}} O_X \ket{\omega_{i}} \bra{\omega_{i}} O_i \ket{\omega_{i}} \right| \nonumber \\
\leq e^{-O(\ell\GapParent)} \norm{O_X} \norm{O_i}.
\end{align}
Choosing $O_i = (A_{i}^{-1})^\dagger A_{i}^{-1}$ to disentangle the PEPS tensor $A_{i}$, 
and dividing by $\norm{A_i^{-1} \ket{\omega_i}}^2=\bra{\omega_i} (A_i^{-1})^\dagger A_i^{-1} \ket{\omega_i}$ gives
\begin{multline}
\left |\frac{\bra{\omega_{i}} O_X \otimes (A_{i}^{-1})^\dagger A_{i}^{-1} \ket{\omega_{i}}}{\bra{\omega_{i}} (A_{i}^{-1})^\dagger A_{i}^{-1} \ket{\omega_{i}}} - \bra{\omega_{i}} O_X \ket{\omega_{i}}  \right| \\
\leq e^{-O(\ell\GapParent)} \norm{O_X} \frac{\norm{(A_{i}^{-1})^\dagger A_{i}^{-1}}}{\bra{\omega_{i}} (A_{i}^{-1})^\dagger A_{i}^{-1} \ket{\omega_{i}}},
\end{multline}
which can be upper bounded by
$ e^{-O(\ell\GapParent)} \norm{O_X} \kappa(A_{i})^2$,
using $\kappa(A)=\kappa(A^{-1})$.
Equipped with this bound, we can iteratively approximate the expectation value $\bra{\omega} O_X \ket{\omega}/\langle\omega\vert\omega\rangle$ using the triangle inequality $O(\ell^{\DimSpatial-1})$ times and bounding the error as we move from $\ket{\omega_N}=\ket{\omega}/\norm{\ket{\omega}}$ to $\ket{\omega_{N-N_b}}\eqqcolon\ket{\omega_*}$:
\begin{equation} \label{eq:LTQO}
	\Bigl|\bra{\omega_*} O_X \ket{\omega_*} - \frac{\bra{\omega} O_X \ket{\omega}}{\langle\omega\vert\omega\rangle} \Bigr|
	\leq \ell^{\DimSpatial-1} e^{-O(\ell\GapParent)} \kappa_*^{2} \norm{O_X} .
\end{equation}
For any given error $\ErrorObservablePEPSToPatch>0$ we can choose a sufficiently large $\ell$ as in eq.\ (\ref{ell}),
defining a patch of the tensor network of radius $\ell$ around the observable $O_X$ that can be disentangled from the rest of the state while not changing the expectation value of $O_X$ by more than~$\ErrorObservablePEPSToPatch$. For constant lattice dimension~$\DimSpatial$, well-conditioned $\kappa_*=O(\poly(N))$, $\ErrorObservablePEPSToPatch=1/O\bigl(\poly(N)\bigr)$, and $\norm{O_X} = O(\poly(N))$, a choice of $\ell = O(\log(N))$ suffices. 
We will now turn to showing how to compute this expectation value in quasi-polynomial time $O(2^{(\log N )^{O(1)}})$.
We write the state vector $\ket{\omega_*}$ as the formally normalised PEPS
$\ket{\omega_*} = 
{A_{N-N_b} \cdot \dots \cdot A_1 \ket{\phi}^{\otimes n}}/{\norm{A_{{N-N_b}} \cdot \dots \cdot A_1 \ket{\phi}^{\otimes n}}}$.
Since we have disentangled all PEPS tensors $A_i$ on a boundary surface, the PEPS $\ket{\omega_*}$ is a tensor product of the patch $\ket{\omega_P}$ and the 
remainder PEPS $\ket{\omega_R}$, $\ket{\omega_*}
    =\ket{\omega_R}\otimes
     \ket{\omega_P} $,
\begin{multline} 
    \ket{\omega_*} =\frac{\bigotimes_{r
                       \in
                        R}
           A_r
           \ket{\phi}^{\otimes
                       n_R}}
          {\norm{\bigotimes_{r
                             \in
                              R}
                 A_r
                 \ket{\phi}^{\otimes
                             n_R}}}\otimes
     \frac{\bigotimes_{p
                       \in
                        P}
           A_p
           \ket{\phi}^{\otimes
                       n_P}}
          {\norm{\bigotimes_{p
                             \in
                              P}
           A_p
           \ket{\phi}^{\otimes
                       n_P}}}.
\label{eq:productstate}											
\end{multline}
Since $O_X$ acts only on $\ket{\omega_P}$ and 
$\langle{\omega_R}\vert{\omega_R}\rangle=1$, one gets
\beq
\bra{\omega_*}O_X\ket{\omega_*}= \bra{\omega_P} O_X \ket{\omega_P} 
= \frac{\bra{\phi} \bigotimes_{p \in P} A_p^\dag O_X A_p \ket{\phi}}{\bra{\phi} \bigotimes_{p \in P} A_p^\dag A_p \ket{\phi}}.
\label{eq:patchcontract}
\eeq
The left tensor factor $\ket{\omega_R}$ in eq.\ (\ref{eq:productstate}) has been reduced to a scalar $1$ by construction in this step.
We have hence removed the need to contract any part of the PEPS outside of the patch or to 
compute a global norm. All remaining computations can be performed locally.
The tensor networks in both the numerator as well as the denominator in eq.\ (\ref{eq:patchcontract}) can be contracted exactly in time $(\DimVirtual\DimPhysical)^{O(\ell^\DimSpatial)}$ by summing over all $2(\DimSpatial\ell^\DimSpatial)$ indices of dimension $\DimVirtual$ and $\ell^\DimSpatial$ indices of 
physical dimension $\DimPhysical$, resulting in 
$(\DimVirtual\DimPhysical)^{O(\ell^\DimSpatial)}$ terms,
for constant $\DimSpatial\geq 2$ and constant $\ell$ in polynomial and for $\ell=O(\log(N))$ in quasi-polynomial time.

\emph{Computation of expectation values on a quantum computer.}
We have seen that the expectation value of a local observable can be approximated by computing the expectation value on a small patch of radius $\ell=O(\log(N))$ instead of the full PEPS. This fact suggests that the desired expectation value could also be computed on a \emph{quantum} computer acting on only $O(\log(N))$ qubits, instead of simulating the full system of size $N$. 
Indeed, this observation contributes to the discussion on feasible applications of  
a small quantum computer consisting of tens or hundreds, but not thousands or more qubits.

\emph{Hardness of tensor network contraction.}
The result obtained here is not in conflict with the known hardness result \cite{Contraction} for contracting general PEPS, an argument
that we lay out in more detail in the supplementary material. 
The key reason is the restriction to PEPS which are ground states of local Hamiltonian with constant spectral gap. 
This assumption significantly reduces the computional power of a PEPS oracle viewed as a computational resource \cite{Contraction}.
   
\emph{Implications for gaps of transfer operators.}
We emphasize that the statement of Theorem \ref{thm:PEPSobservable}
and the content of Conjecture  \ref{conj:strongPEPSconj} imply that the transfer operator is gapped, 
an observation that is often made in practice.  It contributes to providing 
evidence why numerical contraction methods of PEPS perform so well in numerical studies.

\emph{Injective PEPS with uniformly gapped parent Hamiltonians satisfy a variant of local topological quantum order.}
In ref.\ \cite{cirac2013robustness} it is shown that parent Hamiltonians of translationally invariant, injective MPS satisfy the 
\emph{local topological quantum order} (LTQO) condition. Combining eq.\ (\ref{eq:productstate}) and (\ref{eq:LTQO}) yields
\begin{equation*}
	\Bigl|\frac{\bra{\omega_P} O_X \ket{\omega_P}}{\langle\omega_P\vert\omega_P\rangle} - \frac{\bra{\omega} O_X \ket{\omega}}{\langle\omega\vert\omega\rangle} \Bigr|
	\leq \norm{O_X}  \ell^{\DimSpatial-1} e^{-O(\ell\GapParent)} \kappa_*^{2},
\end{equation*}
which might superficially appear to satisfy the LTQO condition in ref.\ \cite[Def.\ 2]{cirac2013robustness}, but actually does not. Rather, the two statements differ in the type of \emph{boundary terms} allowed: while the cited LTQO condition only allows to strictly remove local Hamiltonian terms along a boundary surface, the parent Hamiltonians constructed in our proof actually do add boundary terms to enforce the uniqueness of the ground state. Thus, in this sense our proof does not imply LTQO for injective PEPS as defined, but rather {a variant of LTQO} with unique ground states, which may be of independent interest.

\emph{Conclusion and outlook.}
In this work, we have shown that expectation values of local observables in  
PEPS that naturally approximate ground states of Hamiltonian models can be computed 
in quasi-polynomial time. In this way, we contribute a complexity-theoretic picture to the 
widely observed common observation that in numerical approaches, such computations are well feasible. 
It is key to the argument that for a given local observable, not the entire remainder more than a constant distance away from the support of the relevant observable needs to be disentangled: This would lead to a norm 
that cannot be bounded in a way necessary to invoke the clustering of correlations. The 
techniques that we introduce, however, are powerful enough to arrive at the conclusion of a
quasi-polynomial contraction.
It remains a very interesting problem for future work to extend our results to $G$-injective \textsc{PEPS}.
What we have shown here can to an extent be seen as the ground state analogue of an insight
that has already been established for high-temperature Gibbs states, for which both an
efficient approximation in terms of tensor network states with polynomial bond dimension
and an efficient computation of expectation values have both been identified 
\cite{ThermalPEPOs,Intensive}.
We hence contribute to demystifying the complexity of 
contracting tensor network states, coming to the conclusion that the
situation for higher dimensional systems is not that different compared to 1D systems, for which
the DMRG approach provides the workhorse of numerical studies. Our work can be seen
as a further invitation to the program of capturing condensed matter systems in terms of tensor network
states. 

\emph{Acknowledgements.}
We thank I. Arad, M.\ Kastoryano, Y. Ge, Z. Landau, A. Moln\'{a}r, N.\,Schuch, and F.\,Verstraete for discussions and the ERC (TAQ), the 
EC (SIQS, RAQUEL, AQuS), the Templeton Foundation, 
the DFG (EI 519/7-1, CRC 183), and the A.\ v.\ Humboldt Foundation for support.


%

\bigskip
\section{Appendix}

\subsection*{Proof of Theorem 1}

In this section, we lay out the proof of the main text in more detail, at the expense of some reduncancy.
For an unnormalised, injective PEPS~$\ket{\omega}$, specified by the set of tensors $\{A_i\}$, with a gapped, local parent Hamiltonian $H_*$ \cite{Perez-Garcia-QIC-2008},
define a sequence of Hamiltonians as follows. Let $H_{i}$, with $0 \leq i \leq N$, be a family of parent Hamiltonians on the same $\DimSpatial$-dimensional lattice, with formally \emph{normalised} PEPS ground states 
\begin{equation}
	\ket{\omega_{i}} = \frac{A_{i} \cdot \dots \cdot A_1 \ket{\phi}^{\otimes n}}{\norm{A_{i} 
	\cdot \dots \cdot A_1 \ket{\phi}^{\otimes n}}},
\end{equation}
so that $H_N=H_*$.
Here $n=O(N)$ denotes the number of edges.
Let us assume that the last $N_b=O(\ell^{\DimSpatial-1})$ tensors in the sequence $(A_i)_{0\leq i \leq N}$ constitute a boundary at distance $\ell$ in 
graph theoretical distance on the lattice around observable~$O_X$.
This sequence of parent Hamiltonians can always be constructed for an injective PEPS \cite{PEPS,Perez-Garcia-QIC-2008}. By our assumptions $H_*$ is uniformly gapped by $\Delta_*$, therefore $\Gap(H_{i}) \geq \Delta_*$ for all $H_i$, following in this step precisely the logic of ref.\ \cite{PEPSgen1,PEPSgen2,ge2016rapid}.

Given this sequence of gapped parent Hamiltonians, which are defined on the same lattice as the Hamiltonian~$H_*$, 
the exponential clustering theorem \cite{hastings2004lieb,math-ph/0506030} applies to each Hamiltonian~$H_i$. That is, we have for each $i$ and fixed $O_X$
\begin{align}
\left |\bra{\omega_{i}} O_X \otimes O_i \ket{\omega_{i}} - \bra{\omega_{i}} O_X \ket{\omega_{i}} \bra{\omega_{i}} O_i \ket{\omega_{i}} \right| \nonumber \\
\leq e^{-O(\ell\GapParent)} \norm{O_X} \norm{O_i}.
\end{align}
Let us choose $O_i = (A_{i}^{-1})^\dagger A_{i}^{-1}$ to disentangle the $i$-th PEPS tensor $A_{i}$, that is
\begin{widetext}
\beq
\abs{\bra{\omega_{i}} O_X \otimes (A_{i}^{-1})^\dagger A_{i}^{-1} \ket{\omega_{i}} - \bra{\omega_{i}} O_X \ket{\omega_{i}} \bra{\omega_{i}} (A_{i}^{-1})^\dagger A_{i}^{-1} \ket{\omega_{i}}} \leq e^{-O(\ell\GapParent)} \norm{O_X} \norm{(A_{i}^{-1})^\dagger A_{i}^{-1}}
\eeq
\end{widetext}
and divide by $\norm{A_i^{-1} \ket{\omega_i}}^2=\bra{\omega_i} (A_i^{-1})^\dagger A_i^{-1} \ket{\omega_i}$, yielding
\begin{multline}
\left |\frac{\bra{\omega_{i}} O_X \otimes (A_{i}^{-1})^\dagger A_{i}^{-1} \ket{\omega_{i}}}{\bra{\omega_{i}} (A_{i}^{-1})^\dagger A_{i}^{-1} \ket{\omega_{i}}} - \bra{\omega_{i}} O_X \ket{\omega_{i}}  \right| \\
\leq e^{-O(\ell\GapParent)} \norm{O_X} \frac{\norm{(A_{i}^{-1})^\dagger A_{i}^{-1}}}{\bra{\omega_{i}} (A_{i}^{-1})^\dagger A_{i}^{-1} \ket{\omega_{i}}}.
\end{multline}
We can simplify this to 
\begin{eqnarray}
&|\bra{\omega_{i-1}} O_X &\ket{\omega_{i-1}} - \bra{\omega_{i}} O_X \ket{\omega_{i}}  | \\ \nonumber
&\leq& e^{-O(\ell\GapParent)} \norm{O_X} \frac{\norm{(A_{i}^{-1})^\dagger A_{i}^{-1}}}{\bra{\omega_{i}} (A_{i}^{-1})^\dagger A_{i}^{-1} \ket{\omega_{i}}} \nonumber\\
&\leq& e^{-O(\ell\GapParent)} \norm{O_X} \left( \frac{\sigma_{
\max}(A_{i}^{-1})}{\sigma_{\min} (A_{i}^{-1})} \right)^2 \nonumber\\
&=& e^{-O(\ell\GapParent)} \norm{O_X} \kappa(A_{i})^2
\end{eqnarray}
using $\kappa(A)=\kappa(A^{-1})$.
Let us emphasise that~$\ket{\omega_{i-1}}$ is again a normalised state since we effectively absorbed the norm change from~$A_i^{-1}\ket{\omega_i}$ to~$\ket{\omega_{i-1}}$ into the \emph{local} condition number~$\kappa(A_i)^2$.
Equipped with this bound, we can iteratively approximate the expectation value $\bra{\omega} O_X \ket{\omega}/\langle\omega\vert\omega\rangle$ using the triangle inequality $O(\ell^{\DimSpatial-1})$ times and bound the error as we move from $\ket{\omega_N}=\ket{\omega}/\norm{\ket{\omega}}$ to $\ket{\omega_{N-N_b}}\eqqcolon\ket{\omega_*}$:
\begin{equation} \label{eq:LTQO}
	\Bigl|\bra{\omega_*} O_X \ket{\omega_*} - \frac{\bra{\omega} O_X \ket{\omega}}{\langle\omega\vert\omega\rangle} \Bigr|
	\leq \ell^{\DimSpatial-1} e^{-O(\ell\GapParent)} \kappa_*^{2} \norm{O_X} .
\end{equation}
Thus, for any given error $\ErrorObservablePEPSToPatch>0$ we can choose a sufficiently large
\begin{equation}
	\ell \in O\left(\frac{2 \ln(\kappa_*) + \ln(\ErrorObservablePEPSToPatch^{-1}) + \ln(\norm{O_X})}{\GapParent}\right)
\end{equation}	
which defines a patch of the tensor network of radius $\ell$ around the observable $O_X$ that can be disentangled from the rest of the state while not changing the expectation value of $O_X$ by more than~$\ErrorObservablePEPSToPatch$. For the common case of constant lattice dimension~$\DimSpatial$, well-conditioned $\kappa_*=O(\poly(N))$, $\ErrorObservablePEPSToPatch=1/O\bigl(\poly(N)\bigr)$, and $\norm{O_X} = O(\poly(N))$, a choice of $\ell = O(\log(N))$ suffices. 

We will now show how to compute this expectation value in quasi-polynomial time $O(2^{(\log N )^{O(1)}})$.
Let us write the state vector $\ket{\omega_*}$ as the formally normalised PEPS
\begin{equation}
\ket{\omega_*} = 
\frac{A_{N-N_b} \cdot \dots \cdot A_1 \ket{\phi}^{\otimes n}}{\norm{A_{{N-N_b}} \cdot \dots \cdot A_1 \ket{\phi}^{\otimes n}}}.
\end{equation}
Since we have disentangled all PEPS tensors $A_i$ on a boundary surface, the PEPS $\ket{\omega_*}$ is a tensor product of the patch $\ket{\omega_P}$ and the 
remainder PEPS $\ket{\omega_R}$,
\begin{multline} 
    \ket{\omega_*}
    =\ket{\omega_R}\otimes
     \ket{\omega_P} \\
    =\frac{\bigotimes_{r
                       \in
                        R}
           A_r
           \ket{\phi}^{\otimes
                       n_R}}
          {\norm{\bigotimes_{r
                             \in
                              R}
                 A_r
                 \ket{\phi}^{\otimes
                             n_R}}}\otimes
     \frac{\bigotimes_{p
                       \in
                        P}
           A_p
           \ket{\phi}^{\otimes
                       n_P}}
          {\norm{\bigotimes_{p
                             \in
                              P}
           A_p
           \ket{\phi}^{\otimes
                       n_P}}}.
\label{eq:productstate}											
\end{multline}
Since $O_X$ acts only on $\ket{\omega_P}$ and 
$\langle{\omega_R}\vert{\omega_R}\rangle=1$, one gets
\beq
\bra{\omega_*}O_X\ket{\omega_*}= \bra{\omega_P} O_X \ket{\omega_P} 
= \frac{\bra{\phi} \bigotimes_{p \in P} A_p^\dag O_X A_p \ket{\phi}}{\bra{\phi} \bigotimes_{p \in P} A_p^\dag A_p \ket{\phi}}.
\label{eq:patchcontract}
\eeq
Note that in this step the left tensor factor $\ket{\omega_R}$ in eq.\ (\ref{eq:productstate}) has been reduced to a scalar $1$ by construction.
Thus, we have completely removed the need to contract any part of the PEPS outside of the patch, or to compute a global norm of the PEPS. All remaining computations can be performed locally on the patch.
The tensor networks in both the numerator as well as the denominator in eq.\ (\ref{eq:patchcontract}), can be contracted exactly in time $(\DimVirtual\DimPhysical)^{O(\ell^\DimSpatial)}$ by summing over all $2(\DimSpatial\ell^\DimSpatial)$ indices of dimension $\DimVirtual$ and $\ell^\DimSpatial$ indices of 
physical dimension $\DimPhysical$, resulting in 
$(\DimVirtual\DimPhysical)^{O(\ell^\DimSpatial)}$ terms.
Thus, for constant $\DimSpatial\geq 2$ and constant $\ell$ in polynomial time, and for $\ell=O(\log(N))$ in quasi-polynomial time.

\subsection*{Hardness of tensor network contraction}

Ref.\ \cite{Contraction} derives the hardness of PEPS contraction by showing an equivalence with \pp, using $\pp=\postbqp$ \cite{aaronson2005quantum}. This is achieved by relating the PEPS to a quantum circuit in \emph{measurement based quantum computing}, accepting only outcomes in which 
no Pauli corrections are required. Projections are hence an inherent part of the hardness construction, while
in our argument, invoking injectivity, projections are not considered. 
Ref.\ \cite{Contraction} also considers those injective PEPS achieved through a perturbation of the construction above. Simulating measurements on such a PEPS would remain \pp-hard, and thus such PEPS can only be a ground states of a local Hamiltonian with at most an exponentially small gap (unless \pp$=$\qma, which is considered unlikely \cite{vyalyi2003qma}), while ground states of at least poly-gapped Hamiltonians (or even constant, as in our case) is in \textsf{(F)}\qma. Thus, the complexity of measuring local observables on the set of PEPS ground states of local Hamiltonians with constant spectral gap is of lower complexity (at most \qma, and due to 
Theorem 1
probably strictly lower) than the set of all PEPS, according to standard complexity theoretic assumptions.

The same ref.\ \cite{Contraction} also shows that computing the \emph{norm} of a PEPS is equivalent to computing an expectation value on a general PEPS, and thus hard. Our proof neither assumes normalization of the input PEPS, nor does it run into this issue as we explicitly carry normalization factors through the entire proof, that never need to be evaluated globally, as it turns out. Only in the final step, the normalization of the local patch of poly-logarithmic size has to be computed explicitly.

\subsection*{Discussion of the computation of expectation values on a quantum computer}

Our classical algorithm is to some extent related to methods for preparing PEPS on a quantum computer \cite{PEPSgen1, PEPSgen2, ge2016rapid}. Therefore, it comes as no surprise that under the same assumptions as Theorem 1
we can also prepare the PEPS defined by the patch, and then estimate the local observable. In particular, the method of ref.\ \cite{ge2016rapid} is able to prepare our patch of size $O(\ell^\DimSpatial)$ in time $T=O(\ell^\DimSpatial \text{polylog}(\ell/\varepsilon))$ with error $\varepsilon$ in trace distance. The circuit can also be parallelized to depth $D=O(\text{polylog}(\ell/\varepsilon))$. Using the standard Chernoff bound, it's clear that $O(1/\varepsilon^2)$ independent preparations and measurements suffices to estimate $\langle O_X \rangle$ up to $\varepsilon$ with constant probability of error. Thus, we can approximate the expectation value on a small-scale quantum computer consisting of $\ell^\DimSpatial=O(\log(N))$ spins in $\tilde{O}(\ell^\DimSpatial/\varepsilon^2)$ quantum time. 

\subsection*{Implications for the gap of the transfer operator}

Contracting a higher-dimensional PEPS to a one-dimensional MPS, one finds that the 
injectivity of the MPS is inherited by that of the PEPS. This implies the uniqueness of the 
largest eigenvalue of the one-dimensional transfer operator. From the exponential decay of correlations
one can hence directly derive a lower bound to the gap of the transfer operator, in fact bounded
by the gap of the parent Hamiltonian $H_*$.

In this section, we provide further detail on the implications 
for the gap of the transfer operator. We assume the validity of Conjecture 2 and the assumptions of Theorem 
1, and consider a translationally invariant PEPS. For the entire
PEPS, the transfer operator per site $v\in V$ is given by
\begin{eqnarray}
	E
	&=&\sum_{i
	       =1}^\DimPhysical
	 \sum_{\overset{j_1,\dots, j_r=1}
	               {k_1,\dots, k_r=1}}^\DimVirtual
	 T_{v;i,j_1,\dots, j_r} T^\ast_{v,i,k_1,\dots, k_r} \nonumber\\
	&=& |j_1,\dots, j_r\rangle \langle k_1,\dots, k_r|,
\end{eqnarray}
from which expectation values of local observables can be computed.
For a given one-dimensional line $L\subset V$ of the $\DimSpatial$-dimensional cubic lattice, we can define the resulting one-dimensional transfer operator as
\begin{eqnarray}
	e&=&  {\cal C}_{V\backslash L}\biggl(\prod_{v\in V} E_v\biggr)\\
		&=&   \sum_{i=1}^\DimPhysical \sum_{j_1,j_2,k_1,k_2=1}^\DimVirtual(t_{i,j_1,j_2}  t^\ast_{i,k_1,k_2}
	)
	|j_1,j_2\rangle \langle k_1,k_2| ,\nonumber
\end{eqnarray}
achieved upon contraction of all virtual indices over $V\backslash L$, defining the matrices $\{t_{i,j_1,j_2}:i=1,\dots, \DimPhysical, j_1,j_2=1,\dots, \DimVirtual\}$ 
of an MPS along $L$. Correlation functions between observables $O_A$ and $O_B$
supported on sites $1,x+1\in  L$, respectively,  so that $O_X= O_A \,O_B$,
can then be computed as
\begin{equation}
	\langle O_A\otimes O_B\rangle ={\rm tr} (e_{O_A} e^{x} e_{O_B} e^{|L|-x-2})/ {\rm tr}(e^{|L|}),
	\end{equation}
where for an observable~$O$ at a site we have that $e_O=\sum_{i,j=1}^\DimPhysical \, \sum_{j_1,j_2,k_1,k_2=1}^\DimVirtual
(t_{i,j_1,j_2} O_{i,j} t^\ast_{j,k_1,k_2}) |j_1,j_2\rangle \langle k_1,k_2|$. Invoking the above assumptions,  
there exist constants $c_1,c_2>0$ such that  
\begin{equation}\label{Decay}
	| \langle O_A\otimes O_B\rangle -  \langle O_A\rangle\langle O_B\rangle | \leq c_1 e^{-c_2  x \delta}. 
\end{equation}
	As is well known, choosing without loss of generality the largest eigenvalue of $e$ as $\lambda_1=1$, 
	one has
	$e^x = |r\rangle\langle l| + \sum_{j>1} \lambda_j^x |r_j\rangle\langle l_j| $. Since 
	injectivity of the MPS is inherited from the injectivity of the PEPS, implying 
	that the largest eigenvalue of the transfer
	operator is unique, and since eq.\ (\ref{Decay})  is assumed to hold
for all observables $O_A$ and $O_B$, from
\begin{eqnarray}
	|{\rm tr} (e_{O_A} e^{x} e_{O_B} e^{|L|-x-2}) &-& \langle l| e_{O_A}|r\rangle \langle l |e_{O_B}|r\rangle  | = O(e^{-|L|})\nonumber \\ 
	&+&  \sum_{j>1} \lambda_j^x  
	\langle l| e_{O_A}|r_j\rangle\langle l_j |e_{O_B}| r \rangle  
\end{eqnarray}	
it follows that
\begin{equation}
	\frac{\lambda_2}{\lambda_1}\leq e^{-c_2\delta},
\end{equation}
such that the transfer operator along the line $L$ itself inherits the gap of the parent Hamiltonian.

\end{document}